\documentclass[12pt]{article}
\usepackage{subfloat}
\pdfoutput = 1
\begin{document}
\date{Today}
\title{{\bf{\Large  Analytic study of properties of holographic superconductors away from the probe limit }}}

\author{
{\bf {\normalsize Sunandan Gangopadhyay}$^{a,b}
$\thanks{sunandan.gangopadhyay@gmail.com, sunandan@iucaa.ernet.in, sunandan@bose.res.in}}\\
$^{a}$ {\normalsize Department of Physics, West Bengal State University, Barasat, India}\\
$^{b}${\normalsize Visiting Associate in Inter University Centre for Astronomy \& Astrophysics,}\\
{\normalsize Pune, India}\\[0.1cm]
}
\date{}

\maketitle

\begin{abstract}
{\noindent In this paper, based on the Sturm-Liouville eigenvalue approach,
we analytically investigate the properties of holographic superconductors
in the background of pure Einstein and Gauss-Bonnet gravity taking into account the backreaction of the spacetime.
Higher value of the backreaction parameter results in a harder condensation to form in both cases. The
analytical results obtained are found to be in good agreement with the existing numerical results.  }
\end{abstract}
\vskip 1cm

\section{Introduction}

The correspondence between anti-de Sitter and conformal field theories (AdS/CFT) has been a powerful tool
to analyse strongly coupled quantum field theories. It provides a correspondence between a gravity theory 
in a $(d+1) $ dimensional $AdS$ spacetime and a conformal field theory (CFT) living on its $d$-dimensional 
boundary \cite{adscft1}-\cite{adscft4}. Recently, the correspondence has been used to provide some meaningful
theoretical insights to understand the physics of high $T_c$ superconductors from the gravitational dual.

The central idea behind holographic superconductors comes from the observation that below a critical temperature,
electrically charged black holes become unstable to the formation of scalar hair. 
The mechanism behind this condensation is the breaking of a local $U(1)$ symmetry near 
the event horizon of the black hole \cite{hs1}-\cite{hs6}.
However, the investigations in most cases have been carried out in the ``proble limit" which essentially means that the 
backreaction of the spacetime has been neglected. Backreaction of the spacetime was considered in \cite{hs6}-\cite{hs16} 
for a $2+1$-dimensional holographic superconductor where it was found that 
even an uncharged scalar field can form a condensate. In \cite{betti}, $3+1$-dimensional holographic
superconductors in pure Einstein and Gauss-Bonnet gravity have been studied taking backreaction into account numerically.
This study was motivated by the fact that earlier such studies have been made on black holes in Gauss-Bonnet gravity
in the probe limit \cite{hs8}.
A lot of work has been done thereafter to study the properties of 
holographic superconductors away from the probe limit \cite{gub}-\cite{pan}.

In this paper, we try to substantiate the numerical results of \cite{betti} analytically. We apply the Sturm-Liouville (SL)
method developed in \cite{hs7} to analytically find the relation between the critical temperature and the charge density
both in Einstein gravity and Gauss-Bonnet gravity taking backreaction of the spacetime into account. This method
has been used earlier in the probe limit with considerable success \cite{hs10}-\cite{sgan}. Our analysis would also help in examining
the applicability of the method in the presence of the backreaction.



This paper is organized as follows. In section 2, we provide the basic holographic set up for the holographic superconductors,
considering the background of a $4+1$-dimensional electrically charged black hole in anti-de Sitter spacetime.
In section 3, taking into account the backreaction of the spacetime in Einstein gravity, 
we compute the critical temperature in terms of a solution to the SL eigenvalue problem. In section 4,
we carry out the same analysis in Gauss-Bonnet gravity. We conclude finally in section 5.

\section{Basic set up }
To begin with, we first write down the action for the formation of scalar
hair on an electrically charged black hole in $4+1$-dimensional anti-de Sitter spacetime. This reads
\begin{eqnarray}
S=\frac{1}{16\pi G}\int~d^5 x~\sqrt{-g}\left(R-2\Lambda +\frac{\alpha}{4}(R^{\mu\nu\lambda\rho}R_{\mu\nu\lambda\rho}-4R^{\mu\nu}R_{\mu\nu}+R^2)+16\pi G \mathcal{L}_{matter}\right)
\label{ac1}
\end{eqnarray}
where $\Lambda=-6/L^2$ is the cosmological constant and $\alpha$ is the Gauss-Bonnet coupling parameter.
$\mathcal{L}_{matter}$ denotes the matter Lagrangian density and takes the form
\begin{eqnarray}
\mathcal{L}_{matter}=-\frac{1}{4}F_{MN}F^{MN}-(D_{M}\psi)^{*} D^{M}\psi-m^2 \psi^{*}\psi~;~M,N=0,1,2,3,4
\label{ac2}
\end{eqnarray}
where $F_{MN}=\partial_{M}A_{N}-\partial_{N}A_{M}$ is the field strength tensor and
$D_{M}\psi=\partial_{M}\psi-ieA_{M}\psi$ is the covariant derivative.
The ansatz for the plane-symmetric black hole metric reads
\begin{eqnarray}
ds^2=-f(r)a^2 (r)dt^2+\frac{1}{f(r)}dr^2+\frac{r^2}{L^2}(dx^2+dy^2 +dz^2)~.
\label{m1}
\end{eqnarray}
We now choose the following ansatz for the gauge field and the scalar field \cite{hs6}
\begin{eqnarray}
A_{M}dx^M=\phi(r)dt~,~\psi=\psi(r)
\label{vector}
\end{eqnarray}
so that the black hole possesses only electric charge.

\noindent The equations of motion for the metric and matter fields computed on this ansatz read
\begin{eqnarray}
\label{e02}
f'(r)+2r\frac{f(r)-2r^2 /L^2}{(r^{2}-2\alpha f(r))}
+\gamma\frac{ r^{3}}{2 f(r) a^2 (r)}\nonumber\\
\times\left(\frac{2e^{2}\phi^{2}(r)\psi^{2}(r)+f(r)(2m^2 a^2 (r) \psi^2 (r) + \phi'^{2}(r))+2f^2 (r) a^2 (r)\psi'^2(r)}
{r^{2}-2\alpha f(r) }\right)=0\\
\label{e01}
a'(r)-\gamma\frac{r^3 (e^2\phi^{2}(r)\psi^2 (r) +a^2 (r) f^2 (r) \psi'^{2} (r))}{a(r)f^2 (r) (r^2 - 2\alpha f(r))}=0\\
\label{e1}
\phi''(r)+\left(\frac{3}{r}-\frac{a'(r)}{a(r)}\right)\phi'(r)
-2\frac{e^2 \psi^2 (r)}{f(r)}\phi(r)=0\\
\psi^{''}(r)+\left(\frac{3}{r}+\frac{f'(r)}{f(r)}+\frac{a'(r)}{a(r)}\right)\psi'(r)
+\left(\frac{e^2\phi^{2}(r)}{f^2 (r) a^2 (r)}-\frac{m^2}{f(r)}\right)\psi(r)=0
\label{e2}
\end{eqnarray}
where $\gamma=16\pi G$ and prime denotes derivative with respect to $r$. 
The fact that $\gamma\neq0$ takes into account the backreaction of the spacetime. This limit also allows one to set $e=1$ without
any loss of generality since the rescalings $\psi\rightarrow \psi/e$, $\phi\rightarrow \phi/e$ 
and $\gamma\rightarrow e^2 \gamma$ can be performed \cite{betti}.

\noindent In order to solve the non-linear equations (\ref{e02})-(\ref{e2}), we need
to fix the boundary conditions for $\phi(r)$ and $\psi(r)$ at the black
hole horizon $r=r_+$ (where $f(r=r_+)=0$ with $a(r=r_+)$ finite) and at the spatial infinity
($r\rightarrow\infty$). At the horizon, we require $\phi(r_+)=0$ and $\psi(r_{+})$ to be finite for the matter fields to be regular. 

\noindent Near the boundary of the bulk, we can set $a(r\rightarrow\infty)\rightarrow1$, so that the spacetime becomes a
Reissner-Nordstr\"{o}m-anti-de Sitter black hole. The matter fields there obey \cite{hs8}
\begin{eqnarray}
\label{bound1}
\phi(r)&=&\mu-\frac{\rho}{r^2}\\
\psi(r)&=&\frac{\psi_{-}}{r^{\lambda_{-}}}+\frac{\psi_{+}}{r^{\lambda_{+}}}
\label{bound2}
\end{eqnarray}
where
\begin{eqnarray}
\label{bound1a}
\lambda_{\pm}&=&2\pm\sqrt{4-3(L_{eff}/L)^2}\\
L_{eff}^2 &=&\frac{2\alpha}{1-\sqrt{1-4\alpha/L^2}}\approx L^2 (1-\alpha/L^2 +\mathcal{O}(\alpha^2))~.
\label{bound3a}
\end{eqnarray}
The parameters $\mu$ and $\rho$ are dual to the chemical potential and charge density of the boundary CFT
and choosing $\psi_{-}=0$, $\psi_{+}$ is dual to the expectation value of the condensation operator $J$ at the boundary.

\noindent Under the change of coordinates $z=\frac{r_{+}}{r}$,  the field equations (\ref{e02})-(\ref{e2}) become
\begin{eqnarray}
\label{e1bbb}
f'(z)+\frac{2r_{+}^{2}}{z^3}\frac{(2r_{+}^{2}-z^2 f(z) )}{(r_{+}^{2}-2\alpha z^2 f(z))}
-\gamma\frac{r_{+}^{2}}{2 z^3 a^2 (z) f(z)}\nonumber\\
\times\frac{\{2r_{+}^{2}\phi^{2}(z)\psi^{2}(z)+f(z)(z^4 \phi'^{2}(z)-6r_{+}^{2}a^2 (z)
\psi^{2}(z))+2a^2 (z)f^2 (z) z^4 \psi'^2(z) \}}{(r_{+}^{2}-2\alpha z^2 f(z))}=0\\
\label{e1bb}
a'(z)+\gamma\frac{r_{+}^{2}}{z^3 a(z) f^{2}(z)}\frac{(r_{+}^{2}\phi^{2}(z)\psi^{2}(z)+a^2 (z)f^2 (z)z^4 
\psi'^{2}(z))}{(r_{+}^{2}-2\alpha z^2 f(z))}=0\\
\label{e1aa}
\phi''(z)-\left(\frac{1}{z}+\frac{a'(z)}{a(z)}\right)\phi'(z) -\frac{2r_{+}^{2}\psi^2 (z)}{z^4 f(z)}\phi(z) =0\\
\psi''(z)-\left(\frac{1}{z}-\frac{a'(z)}{a(z)}-\frac{f'(z)}{f(z)}\right)\psi'(z)
+\frac{r_{+}^{2}}{z^4}\left(\frac{\phi^{2}(z)}{f^2 (z) a^2 (z)}+\frac{3}{f(z)}\right)\psi(z)=0
\label{e1a}
\end{eqnarray}
where prime now denotes derivative with respect to $z$. These equations are to be solved in the
interval $(0, 1)$, where $z=1$ is the horizon and $z=0$ is the boundary.
The boundary condition $\phi(r_+)=0$ now becomes $\phi(z=1)=0$.





\subsection{Effect of backreaction in Einstein gravity}
With the above set up in place, we now move on to investigate the relation between the critical temperature and the charge density. 

\noindent At the critical temperature $T_c$, $\psi=0$, so eq.(\ref{e1bb}) reduces to 
\begin{eqnarray}
a'(z)=0~.
\label{bk1}
\end{eqnarray}
Hence eq.(\ref{e1bbb}) (with $\alpha=0$) and the field equation (\ref{e1aa}) reduces to
\begin{eqnarray}
\label{metric1}
f'(z)+\frac{2}{z^3}(2r_{+(c)}^{2}-z^2 f(z))
-\gamma\frac{z\phi'^{2}(z)}{2}&=&0\\
\phi''(z)-\frac{1}{z}\phi'(z)&=&0.
\label{e1b}
\end{eqnarray}
With the boundary condition (\ref{bound1}), the solution of eq.(\ref{e1b}) reads
\begin{eqnarray}
\phi(z)=\lambda r_{+(c)}(1-z^2)
\label{sol}
\end{eqnarray}
where 
\begin{eqnarray}
\lambda=\frac{\rho}{r_{+(c)}^3}~.
\label{lam}
\end{eqnarray}
This leads to the following solution for the metric from eq.(\ref{metric1}) consistent with the condition $f(z=1)=0$
\begin{eqnarray}
f(z)&=&r_{+(c)}^{2}\left\{\frac{1}{z^2}-(1+\gamma\lambda^2)+\gamma\lambda^2 z^4\right\}=\frac{r_{+(c)}^{2}}{z^2}g_{0}(z)
\label{metr2}
\end{eqnarray} 
where 
\begin{eqnarray}
g_{0}(z)=1-(1+\gamma\lambda^2)z^4+\gamma\lambda^2 z^6 ~.
\label{metr33}
\end{eqnarray} 

\noindent Now using the solution (\ref{sol}), we find that as $T\rightarrow T_c$, 
the equation for the field $\psi$ approaches the limit 
\begin{eqnarray}
-\psi''(z)+\left(\frac{1}{z}-\frac{f'(z)}{f(z)}\right)\psi'(z)-\frac{3}{z^2 g_{0}(z)}\psi(z)
&=&\lambda^2 \frac{(1-z^2)^2}{g_{0}^2 (z)}\psi(z).
\label{e001}
\end{eqnarray}
Near the boundary, we define \cite{hs7}
\begin{eqnarray}
\psi(z)\sim z^3 F(z)
\label{sol1}
\end{eqnarray}
where $F(0)=1$.
Substituting this form of $\psi(z)$ in eq.(\ref{e001}), we obtain
\begin{eqnarray}
- F''(z) + \left(\frac{1}{z}-\frac{f'(z)}{f(z)}-\frac{6}{z}\right)F'(z) + \frac{3}{z}\left\{\left(\frac{1}{z}
-\frac{f'(z)}{f(z)}\right)-\frac{2}{z}
-\frac{1}{z g_{0}(z)}\right\}F(z)=\lambda^2 \frac{(1-z^2)^2}{g_{0}^2 (z)}F(z) \nonumber\\
\label{eq5b}
\end{eqnarray}
to be solved subject to the boundary condition $F' (0)=0$. 

\noindent The above equation can be put in the Sturm-Liouville form 
\begin{eqnarray}
\frac{d}{dz}\left\{p(z)F'(z)\right\}-q(z)F(z)+\lambda r(z)F(z)=0
\label{sturm}
\end{eqnarray}
with 
\begin{eqnarray}
p(z)&=&z^{3}g_{0}(z)\nonumber\\
q(z)&=&3z^5 \{3(1+\gamma \lambda^2)-5\gamma\lambda^2 z^2\}\nonumber\\
r(z)&=&\frac{z^{3}(1-z^2)^2}{g_{0}(z)}~. 
\label{i1}
\end{eqnarray}
With the above identification, we can once again write down the eigenvalue $\lambda^2$ which minimizes the expression 
\begin{eqnarray}
\lambda^2 &=& \frac{\int_0^1 dz\ \{p(z)[F'(z)]^2 + q(z)[F(z)]^2 \} }
{\int_0^1 dz \ r(z)[F(z)]^2}\nonumber\\
&=&\frac{\int_0^1 dz\ z^{3}[g_{0}(z)[F'(z)]^2 
+ 3z^{2}\{3(1+\gamma \lambda^2)-5\gamma\lambda^2 z^2\}[F(z)]^2 ] }
{\int_0^1 dz \ \frac{z^{3}(1-z^2)^2}{g_{0}(z)}[F(z)]^2}~.\nonumber\\
\label{eq5abc}
\end{eqnarray}
To estimate it, we use the following trial function
\begin{eqnarray}
F= F_{\tilde\alpha} (z) \equiv 1 - \tilde\alpha z^2
\label{eq50}
\end{eqnarray}
which satisfies the conditions $F(0)=1$ and $F'(0)=0$.

\noindent Hence, we obtain (with the backreaction parameter $\gamma=0$)
\begin{eqnarray}
\lambda_{\tilde\alpha}^2 = \frac{2(18-27\tilde\alpha+14{\tilde\alpha}^2 )}{6(3-4\ln2)+16(2-3\ln2)\tilde\alpha + (17-24\ln2)
{\tilde\alpha}^2} 
\label{est2}
\end{eqnarray}
which attains its minimum at $\tilde\alpha \approx 0.7218$. 
The critical temperature therefore reads 
\begin{eqnarray}
T_c &=& \frac{1}{4\pi} f'(r_{+(c)}) 
=\frac{1}{\pi\lambda_{\tilde\alpha=0.7218}^{1/3}}\rho^{1/3}\approx 0.196\sqrt\rho 
\label{eqTc}
\end{eqnarray}
which is in very good agreement with the exact $T_c = 0.197\rho^{1/3}$ \cite{hs16}.

\noindent Now in order to include the effect of backreaction, we set $\gamma=0.025$ and put the value of $\lambda^2$
obtained for the corresponding value of $\gamma$ (which in this case is $\gamma=0$) in the right hand side
of eq.(\ref{eq5abc}) to get the  value of $\lambda^2$ for $\gamma=0.025$ 
\begin{eqnarray}
\lambda_{\tilde\alpha}^2 = \frac{1.32909-1.90819\tilde\alpha +0.97677{\tilde\alpha}^2}
{0.06168-0.05909\tilde\alpha + 0.017301{\tilde\alpha}^2} 
\label{est1}
\end{eqnarray}
which attains its minimum at $\tilde\alpha \approx 0.6780$. 
The critical temperature therefore reads 
\begin{eqnarray}
T_c &=& \frac{1}{4\pi} f'(r_{+(c)}) = \frac{1}{\pi}\left(1-\frac{1}{2}\gamma\lambda_{\tilde\alpha=0.7218}^2\right)r_{+(c)}\nonumber\\ 
&=&\frac{1}{\pi}\frac{(1-\frac{1}{2}\gamma\lambda_{\tilde\alpha=0.7218}^2)}{\lambda_{\tilde\alpha=0.6780}^{1/3}}\rho^{1/3}\approx 0.1588\sqrt\rho 
\label{eqTc}
\end{eqnarray}
which is in very good agreement with the exact $T_c = 0.161\rho^{1/3}$ \cite{betti}. Increasing the value of $\gamma$
in steps of $0.025$ and repeating the above process, we can obtain the values of $\lambda^2$ for various values of
the backreaction parameter.

\noindent In the table below, we compare our analytical values obtained by the SL approach with the existing numerical results in the literature \cite{betti}.
\begin{table}[htb]
\caption{A comparison of the analytical and numerical results for the critical temperature and the charge density with backreaction
in Einstein gravity}   
\centering                          
\begin{tabular}{c c c c c c c}            
\hline\hline                        
$\gamma$& $\lambda^{2}_{SL}$  & $(T_{c}/\rho^{1/3})|_{SL}$ &$(T_{c}/\rho^{1/3})|_{numerical}$  \\ [0.05ex]
\hline
0 & 18.23  &   0.196 & 0.197    \\ [0.05ex]  
\hline 
0.025 & 16.3817 &   0.159 & 0.161    \\ [0.05ex] 
\hline 
0.05 & 14.8114 & 0.128 & 0.128    \\ [0.05ex]     
\hline 
0.075 & 13.4269   & 0.103 & 0.089    \\ [0.05ex]  
\hline 
0.1 & 12.1816 & 0.082 & 0.079    \\ [0.05ex]  
\hline 
0.125 & 11.0604 & 0.066 & 0.053    \\ [0.05ex]  
\hline 
0.15 & 10.0647 & 0.053 & 0.045    \\ [0.05ex]    
\hline 
0.175 & 9.1912 & 0.043 &  0.031   \\ [0.05ex]    
\hline 
0.2 & 8.4294 & 0.035 & 0.026    \\ [0.05ex]                 
\end{tabular}\label{E1}  
\end{table}


\subsection{Effect of backreaction in Gauss-Bonnet gravity}
In this section, we study the relation between the critical temperature and the charge density
taking into account the effect of the Gauss-Bonnet coupling parameter $\alpha$. 

\noindent In this case, using eq.(\ref{bk1}), eq.(\ref{e1bbb}) (with $\alpha\neq0$) reduces to
\begin{eqnarray}
f'(z)+\frac{2r_{+(c)}^{2}}{z^3}\frac{(2r_{+(c)}^{2}-z^2 f(z) )}{(r_{+(c)}^{2}-2\alpha z^2 f(z))}
-\gamma\frac{r_{+(c)}^{2}}{2}\frac{z \phi'^{2}(z)}{(r_{+(c)}^{2}-2\alpha z^2 f(z))}=0~.
\label{e10bbb}
\end{eqnarray}
The solution of the above equation upto first order in the Gauss-Bonnet coupling parameter $\alpha$ reads
\begin{eqnarray}
f(z)&=&\frac{r_{+(c)}^{2}}{z^2}\left\{g_{0}(z)+\alpha g_{1}(z)\right\}
\label{metr20z}
\end{eqnarray} 
where 
\begin{eqnarray}
g_{1}(z)=1-2(1+\gamma\lambda^2)z^4+2\gamma\lambda^2 z^6 +(1+\gamma\lambda^2)^2 z^8 -2(1+\gamma\lambda^2)\gamma\lambda^2 z^{10} +\gamma^2 \lambda^4 z^{12}~.
\label{metr33}
\end{eqnarray} 
For $\alpha\neq0$, we define near the boundary
\begin{eqnarray}
\psi(z)\sim z^{\Delta_{+}} F(z)~.
\label{sol1a}
\end{eqnarray}
Substituting this form of $\psi(z)$ in eq.(\ref{e001}), we obtain
\begin{eqnarray}
- F''(z) + \left(\frac{1}{z}-\frac{f'(z)}{f(z)}-\frac{2\Delta_{+}}{z}\right)F'(z) + \frac{\Delta_{+}}{z}\left\{\left(\frac{1}{z}
-\frac{f'(z)}{f(z)}\right)-\frac{\Delta_{+}(\Delta_{+}-1)}{z^2}
-\frac{3}{z^2 (g_0 +\alpha g_1)}\right\}F(z)\nonumber\\
=\lambda^2 \frac{(1-z^2)^2}{(g_0 +\alpha g_1)^2}F(z) \nonumber\\
\label{eq5ba}
\end{eqnarray}
to be solved subject to the boundary condition $F' (0)=0$. 

\noindent The above equation can once again be put in the Sturm-Liouville form 
with 
\begin{eqnarray}
p(z)&=&z^{2\Delta_{+}-3}(g_0 +\alpha g_1)\nonumber\\
q(z)&=&z^{2\Delta_{+}-5}\{\Delta_{+}(2(g_0 +\alpha g_1)-z(g'_{0} +\alpha g'_{1}))
-\Delta_{+}(\Delta_{+}-2)(g_0 +\alpha g_1)-3\} \nonumber\\
r(z)&=&\frac{z^{3}(1-z^2)^2}{g_0 (z)}~. 
\label{i1aa}
\end{eqnarray}
With the above identification, we can once again proceed to find the minimum value of the eigenvalue $\lambda^2$ as
in the earlier section. 

\noindent To estimate it, we first set $\alpha=0.1$, $\gamma=0$ and again use the trial function (\ref{eq50}) to obtain
\begin{eqnarray}
\lambda_{\tilde\alpha}^2 = \frac{1.74-2.5\tilde{\alpha}+1.26194{\tilde\alpha}^2 }{0.0527-0.0496\tilde\alpha + 0.0144{\tilde\alpha}^2} 
\end{eqnarray}
which attains its minimum at $\tilde\alpha \approx 0.7078$. 
The critical temperature therefore reads 
\begin{eqnarray}
T_c &=& \frac{1}{4\pi} f'(r_{+(c)}) 
=\frac{1}{\pi\lambda_{\tilde\alpha=0.7078}^{1/3}}\rho^{1/3}\approx 0.1867\sqrt\rho 
\label{eqTc1a}
\end{eqnarray}
which is in very good agreement with the exact $T_c = 0.185\rho^{1/3}$ \cite{hs8}.

\noindent For $\alpha=0.1$, $\gamma=0.1$, we get
\begin{eqnarray}
\lambda_\alpha^2 = \frac{1.1838-1.4036\tilde\alpha +0.6594{\tilde\alpha}^2}{0.0726-0.0763\tilde\alpha + 0.0239{\tilde\alpha}^2} 
\label{100a}
\end{eqnarray}
which attains its minimum at $\tilde\alpha \approx 0.3495$. In computing this result, we have
used the value of $\lambda^2$ corresponding to $\gamma=0.075$ (with $\alpha=0$) from Table 1 in eq.(\ref{i1aa}) 
to calculate the expression which minimizes $\lambda^2$ and obtained the value of $\lambda^2$ 
corresponding to $\alpha=0.1$, $\gamma=0.1$.
The critical temperature therefore reads 
\begin{eqnarray}
T_c &=& \frac{1}{4\pi} f'(r_{+(c)}) = \frac{1}{\pi}\left(1-\frac{1}{2}\gamma\lambda_{\gamma=0.075, \alpha=0}^2 \right)r_{+(c)}\nonumber\\ 
&=&\frac{1}{\pi}\frac{(1-\frac{1}{2}\gamma\lambda_{\gamma=0.075, \alpha=0}^2)}{\lambda_{\tilde\alpha=0.3495, \alpha=0.1}^{1/3}}\rho^{1/3}\approx 0.066\sqrt\rho 
\label{eqTc}
\end{eqnarray}
which is in very good agreement with the exact $T_c = 0.051\rho^{1/3}$ \cite{betti}.

\noindent In the table below, we compare our analytical values obtained by the SL approach with the existing numerical results in the literature \cite{betti}.

\begin{table}[htb]
\caption{A comparison of the analytical and numerical results for the critical temperature and the charge density with backreaction
in Gauss-Bonnet gravity ($\alpha=0.1$) }   
\centering                          
\begin{tabular}{c c c c c c c}            
\hline\hline                        
$\gamma$  & $(T_{c}/\rho^{1/3})|_{SL}$ &$(T_{c}/\rho^{1/3})|_{numerical}$  \\ [0.05ex]
\hline
0 & 0.1867 & 0.185    \\ [0.05ex]  
\hline 
0.1 & 0.066 & 0.151    \\ [0.05ex] 
\hline 
0.2 & 0.0174 & 0.008    \\ [0.05ex]     
\end{tabular}\label{E1}  
\end{table}


\section{Conclusions}
In this paper, we perform analytic computation of $3+1$-dimensional holographic superconductors in the background
of pure Einstein and Gauss-Bonnet gravity taking into account the backreaction of the spacetime. 
We apply the the Sturm-Liouville eigenvalue problem to obtain the relation between the critical temperature and the charge density 
in both Einstein and Gauss-Bonnet gravity. It is observed that higher value of the backreaction parameter 
results in a harder condensation to form in both cases. Further, the condensation is even harder to form in the presence
of the Gauss-Bonnet parameter. Our results are in very good agreement with the existing numerical results \cite{betti}.

\section*{Acknowledgments} I would like to thank Inter University Centre for Astronomy $\&$ Astrophysics, Pune, India
for providing facilities.


\end{document}